# Self-powered InP Nanowire Photodetector for Single Photon Level Detection at Room Temperature

*Yi Zhu[1, †], Vidur Raj[1, †, *], Ziyuan Li[1], Hark Hoe Tan[1,2], Chennupati Jagadish[1,2] and Lan Fu[1,2,*]*

[1]Department of Electronic Materials Engineering, Research School of Physics and Engineering, The Australian National University, Canberra, ACT 2601, Australia

[2]Australian Research Council Centre of Excellence for Transformative Meta-Optical Systems, Research School of Physics, The Australian National University, Canberra, ACT 2601, Australia

[†] These authors contributed equally to this work.

* To whom correspondence should be addressed: Vidur Raj (vidur.raj@anu.edu.au) and Lan Fu (lan.fu@anu.edu.au)

## Abstract

**Highly sensitive photodetectors with single photon level detection is one of the key components to a range of emerging technologies, in particular the ever-growing field of optical communication, remote sensing, and quantum computing. Currently, most of the single-photon detection technologies require external biasing at high voltages and/or cooling to low temperatures, posing great limitations for wider applications. Here, we demonstrate InP nanowire array photodetectors that can achieve single-photon level light detection at room temperature without an external bias. We use top-down etched, heavily doped p-type InP nanowires and n-type AZO/ZnO carrier selective contact to form a radial p-n junction with a built-in electric field exceeding $3 \times 10^5$ V/cm at 0 V. The device exhibits broadband light sensitivity and can distinguish a single photon per pulse from the dark noise at 0 V, enabled by its design to realize near-ideal broadband absorption, extremely low dark current, and highly efficient charge carrier separation. Meanwhile, the bandwidth of the device reaches above 600 MHz with a timing jitter of 538 ps. The**

proposed device design provides a new pathway towards low-cost, high-sensitivity, self-powered photodetectors for numerous future applications.

## 1. Introduction

Photodetectors that operate at room temperature and without an external bias are of great utility, in particular, if they can achieve single-photon level sensitivity.[1] At present, photomultipliers,[2] superconducting single photon detectors,[3] transition edge sensors,[4] and avalanche photodiodes,[5] remain state-of-art for measuring single photons. However, all of these technologies require bulky operational set-up or external biasing and/or cooling to operate, which limits their usage, especially in applications where long-term external powering or cooling is either infeasible or too complicated due to practical limitations, in applications such as remote sensing, space exploration, wireless environmental monitoring, and medical diagnostics.

Low dimensional III-V compound semiconductor nanostructures have shown tremendous potential for developing high-performance optoelectronic devices such as solar cells, LEDs, lasers, photodetectors, and single-photon emitters.[6] In particular, III-V nanowires, because of their intriguing electrical and optical properties, as well as device design flexibility, have been widely investigated as a replacement for planar devices.[7] Nanowires have also shown promise to detect single photons at room temperature.[8, 9] Recently, Luo et al. demonstrated a room-temperature photon number-resolving detector by using a field-effect transistor configuration in a single CdTe core/shell-like nanowire.[10] Whereas Gibson et al. reported an axial junction p-i-n InP nanowire single photon avalanche photodetector, which operates at room temperature, and can distinguish single photon with very high speed and record low timing jitter.[8] However, so far, to the best of our knowledge, a photodetector that enables single-photon level light detections at room temperature without an external bias has not been reported.

For most optoelectronic devices, including those based on nanowires, the p-n junction serves as a fundamental building block and affects their functionality and performance. Currently, the most common method to achieve either axial or radial p-n junctions in III-V nanowires is through the bottom-up epitaxial growth,[11] which relies on sophisticated and costly growth techniques such as metal-organic chemical vapor deposition (MOCVD) or molecular beam epitaxy (MBE). Although the nanowires synthesized by epitaxial growth tend to have high-quality crystal structures and uniform morphologies, they suffer from doping limitations, and the controlled formation of the p-n junction is highly challenging due to the complex issues related to dopant incorporation, activation, and diffusion.[12, 13] The problem is further complicated if the p-n junction has to be formed radially. Large junction area, unwanted dopant interdiffusion across the radial p-n junction,[13] and segregation of dopants at the radial interface can lead to increased carrier scattering and non-radiative recombination in radial junction device,[14] thereby compromising their distinct advantage (in comparison to axial junction) in simultaneously achieving large active region and efficient charge carrier separation for both light-emitting and detecting devices.[14, 15]

To overcome the challenges in nanowire doping and to form a controlled p-n junction, carrier-selective contacts have been proposed in the field of photovoltaics.[16, 17] In addition to forming a p-n junction and providing carrier selectivity, they can also work as the uniform passivation layers, making them ideal for application in nanowire devices.[16] Other advantages, such as no lattice matching requirement and no dopant requirement for doping, allow for their deposition on a wide range of substrates to form p-n junctions using relatively low-cost deposition techniques such as sputter and atomic layer deposition (ALD). Additionally, most carrier selective contacts are wide bandgap materials with a bandgap > 3 eV, making them transparent in visible and infrared regions, which reduces current loss through parasitic absorption at the operating wavelength.

Here, we demonstrate a high sensitivity, self-powered, non-epitaxial InP nanowire photodetector that enables single-photon level detection at room temperature. We show that through the deposition of AZO/ZnO carrier selective contact on heavily p-type doped InP nanowires, a high built-in electric field of more than $10^5$ V/cm is achieved at 0 V. At the same time, the diameter of AZO/ZnO/InP nanowire and array pitch size has been optimized to achieve near 100% absorption across a broad spectral regime from 350 to 850 nm for optimum device performance. In addition to single-photon level detection, the detector also demonstrates a high bandwidth of above 600 MHz with a timing jitter of 538 ps. We discuss the important technological implications of these devices, which will pave the way toward achieving low-cost nanowire-based self-powered, high sensitivity photodetectors for future applications.

## 2. Results and Discussion

The nanowire arrays were fabricated by the top-down method through inductively coupled plasma (ICP) etching of a $p^+$ InP wafer. The ZnO and AZO layers were deposited sequentially onto nanowires to form the p-n junction. The detailed fabrication steps are presented in the experimental section. The cross-section structure of the fabricated nanowire photodetector, shown schematically in Figure 1a, consists of the Zn/Au back contact, a $p^+$- InP core, a heavily doped n-type ZnO/AZO shell, Au front contact, and a ~300 nm thick layer of SU-8 to isolate the radial heterojunction from the InP substrate. Figure 1b shows the cross-sectional field emission scanning electron microscope (FESEM) image of the bare InP nanowires embedded in 300 nm thick SU-8 at the bottom, which acts as a dielectric layer to separate the n-type contact on the nanowires from p-type contact on the substrate. Figure 1c shows the InP nanowires after a conformal coating of 100 nm AZO/ZnO (90 nm / 10 nm) layers. The ZnO layer acts as an n-type electron selective contact, whereas AZO acts as a transparent conducting oxide, facilitating the lateral transport of carriers to metal contacts.

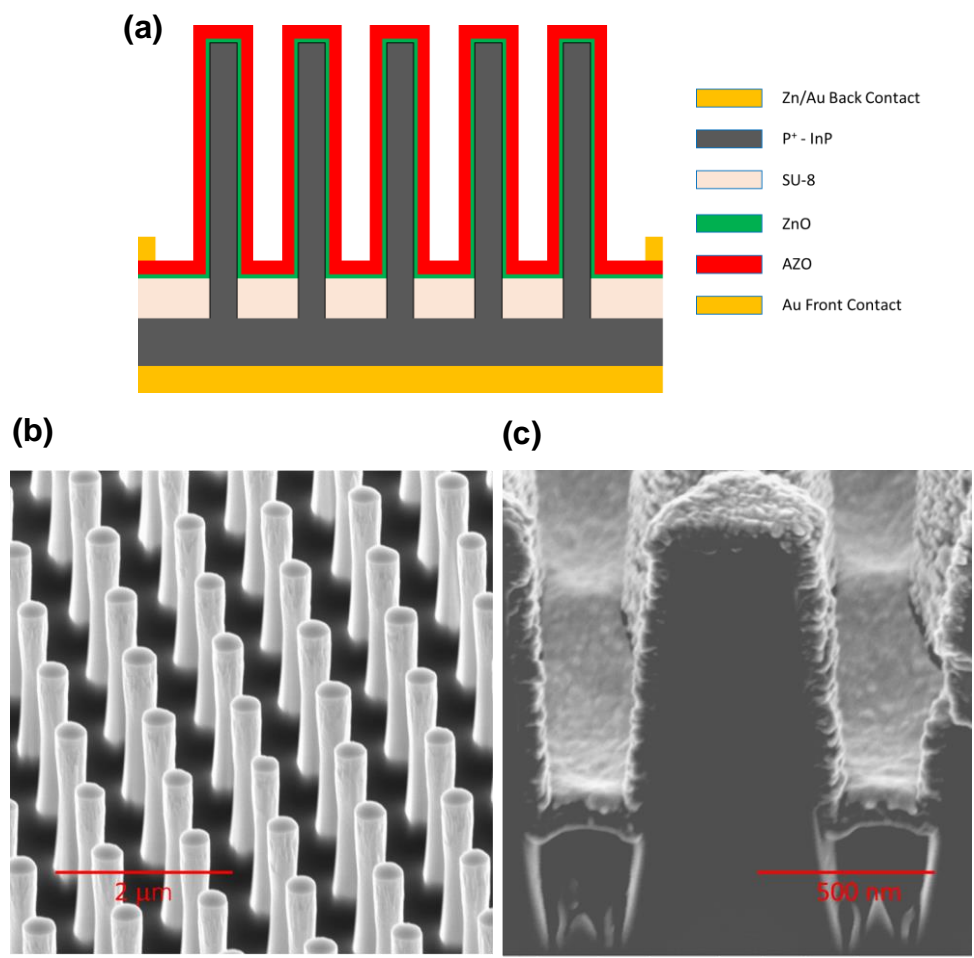

**Figure 1 | Device structure of the InP nanowire photodetector. a** Schematic representation of InP nanowire photodetectors, containing different layers to form a radial p-n junction. **b**

SEM images of InP nanowires embedded in ~300 nm thick SU-8. **c** Cross-section SEM image of a single InP nanowire conformally coated with AZO and ZnO.

## 2.1. *Device design for large broadband optical absorption and ultra-high built-in electric field*

One of the primary requirements for broadband spectral sensitivity in a photodetector is to achieve high absorption within the defined broadband spectral regime. Compared to their thin-film counterparts, nanowire arrays are known to have significantly higher absorption, making them an ideal candidate for highly efficient solar cells and photodetectors. As revealed by previous studies, high absorption in a nanowire is due to an "optical antenna effect",[18, 19] which leads to guided optical modes excitation, and when arranged in an array, these modes become leaky and can start interacting with each other to enhance absorption in a broadband spectral regime.[19, 20] However, to obtain maximum broadband absorption, nanowire pitch and radius must be carefully optimized so that there is maximum overlap between the absorption cross-section of the individual nanowires of the array.[21] Unfortunately, even with the best optimization, an InP nanowire array with a single diameter and fixed pitch is insufficient to obtain 100% absorption across a wide spectral range because the optical resonant modes are highly dependent on nanowire diameter.[19] We recently showed that this limitation could be overcome by utilizing an optimized shell of a non-absorbing oxide layer over InP nanowires.[18] This shell increases the scattering cross-section and reduces the electric-field screening, which maximizes the supported absorbing modes and thus effectively leads to near-ideal wide-spectrum absorption.[18] Through a thorough optimization (for details, see references[18]) of the nanowire array pitch and radius along with AZO/ZnO thickness, we could achieve near 100% absorption between 350 – 850 nm wavelength regime, as shown in Figure 2a.

Moreover, to form the radial p-n junction, we deposit a heavily n-type doped AZO/ZnO over the top-down etched p$^+$-InP nanowire using atomic layer deposition with AZO acting as a transparent conductor and ZnO acting as an electron selective contact. As shown in Figure 2b, a heavily n-type doped ZnO layer forms a type-II heterojunction with p$^+$-InP ($2\times10^{18}$ cm$^{-3}$) with a large valence band offset, and small conduction band offset while also forcing the interface between InP and ZnO to become electron-rich. We also find that ZnO/InP can form effective type-II heterojunction for p-type InP doping of more than $5\times10^{17}$ cm$^{-3}$, however, higher doping was preferred to achieve a high-built-in electric field as will be discussed later. Thereby, high electron selectivity is ensured by impeding holes flow from InP to ZnO due to an extremely low hole conductivity in ZnO and large valence band offset. At the same time, the combined effect of electron accumulation at the interface and electron selectivity of ZnO significantly reduces the interface recombination by screening holes away from the interface,[22] eliminating the requirement of an additional passivation layer.

The most critical aspect of our photodetector device design is that it can achieve a built-in electric field of more than $10^5$ V/cm at 0 V with a short depletion region width of ~ 40 nm. Figures 2c and d respectively show the simulated built-in electric field profile across AZO/ZnO and p$^+$-InP heterojunction, based on the carrier concentrations for AZO, ZnO, and InP of the order of $1x10^{20}$ cm$^{-3}$, $1x10^{19}$ cm$^{-3}$, and 2-3 x $10^{18}$ cm$^{-3}$, respectively, as measured by a four-point probe hall measurement (for more detail see reference [23]). Such a high built-in electric field and a short junction width are a result of the p-n junction formed between the two heavily doped materials and are essential for achieving a single photon photo response at 0 V in our core-shell NW photodetector. The high built-in electric field across the p-n junction is also evidenced by the low breakdown voltage as discussed in section 2.2.

As shown in Figure S1 of the supplementary information, the dark current of the device decreased significantly with increased NW doping, owing to the high built-in electric field and short junction width, which substantially reduces the recombination current in the depletion region. Moreover, it is also important that the high built-in electric field further facilitates efficient charge carrier collection in the highly doped, top-down etched NWs, which have large surface recombination velocity and low minority carrier lifetimes (~50 ps). Both factors are critical for achieving a high signal to noise ratio to enable single photon level detection in our detector. Our result is consistent with previously reported self-powered nanowire photodetectors, where a high built-in electric field led to improved performance[24, 25] and self-powered photodetection.[25]

In general, obtaining such a high built-in electric field at 0 V with low interface recombination would be extremely challenging for an axial or radial III-V epitaxial homo- or heterojunction because of several growth- and fabrication-related complexities, making our photodetector device design unique. It is well known that getting a purely uniform radial growth with high doping concentration is in itself a challenging task when the nanowire has to be grown epitaxially.[26] Although researchers have reported both n- and p-type dopings exceeding $10^{18}$ in III-V nanowires,[27] so far, to the best of our knowledge, there is no report on p$^+$/n$^+$ heterojunction, most probably, because of ineffective dopant incorporation and inter-diffusion of dopants, which become very critical at the nanoscale. On the other hand, because we start the fabrication with a heavily doped p-type InP substrate and proceed with further fabrication at a temperature less than 250 ℃, there is no cross-diffusion of dopants. Also, both AZO and ZnO can have n-type doping density exceeding $10^{20}$ cm⁻³ at a deposition temperature less than 250 ℃ without the requirement of dopants because of donor-like native defects. Given that in our case, both core and shell can easily be heavily doped, the proposed device can achieve such a high built-in electric field.

As a combined result of optical and electrical device design optimization, we achieved a very high quantum efficiency across a broad wavelength regime, as shown in Figure 2e (the extended range EQE data can be found in the supporting information Figure S3). The photodetector had nearly 80% external quantum efficiency (EQE) for most of the measured wavelength, with more than 90% quantum efficiency between 500-700 nm. Even though the absorption is near-ideal, the EQE still suffers from small losses, especially in short (350-450 nm) and long-wavelength (> 900 nm) regimes, mainly due to interface recombination and can be explained by discussing the position-dependent absorption of different wavelength along the length of the NW. In the NW detector, the shorter wavelength light is absorbed in the top segment of the nanowire, so the loss in the EQE curve at shorter wavelengths can be ascribed to the recombination in the top section. More specifically, because the top section of the nanowire has a larger interface than the rest, the ZnO/InP recombination is more pronounced in the shorter wavelength regime. Similarly, the loss at a longer wavelength probably stems from the recombination near the SU-8 region of the nanowire where no ZnO is present such that passivation is relatively poor. To show that the high EQE of our device is highly reproducible and follows a similar trend, we have added measured EQE from another device fabricated in a separate set of experiments in Figure S3.

The corresponding spectral response can be calculated from EQE using equation 1, as follows:

$$pectral\ Response\ (SR) = \frac{q.\lambda}{h.c} \times EQE \qquad (1)$$

In the above equation, EQE is the external quantum efficiency, $q$ is the electronic charge, $h$ is Planck's constant, $c$ and $\lambda$ respectively are the speed and wavelength of light. The *SR* calculated using the equation is plotted in Figure 2f alongside the spectral response of an ideal InP photodetector. To calculate the ideal spectral response the quantum efficiency was assumed to be 100%. Evidently, the photoresponse of our photodetector tracks well with that of an ideal photodetector.

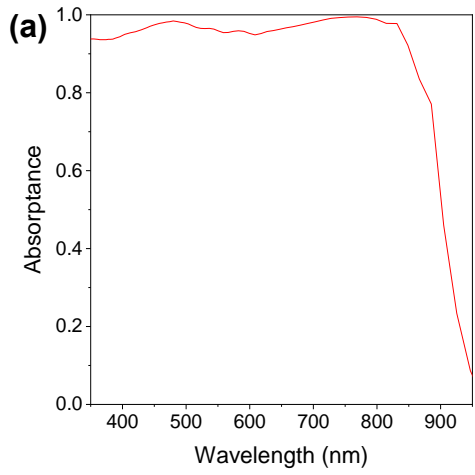

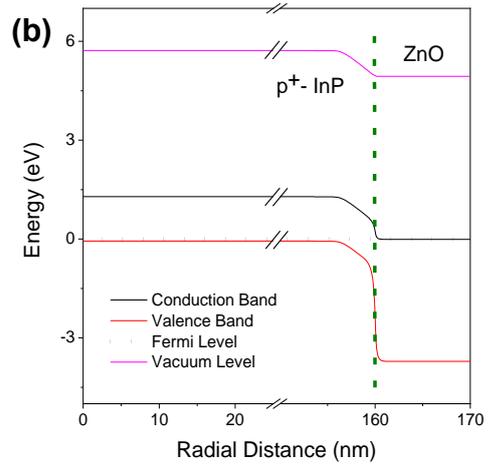

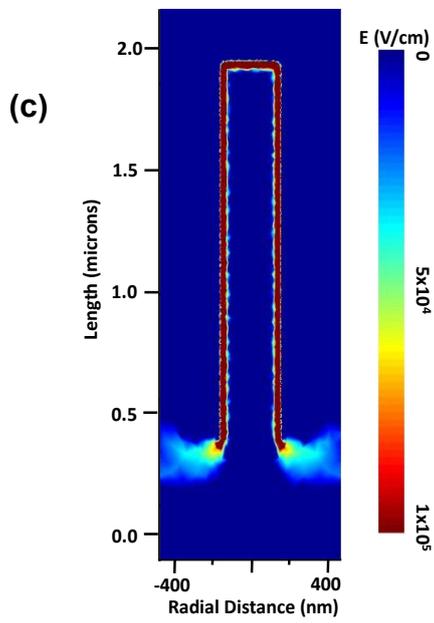

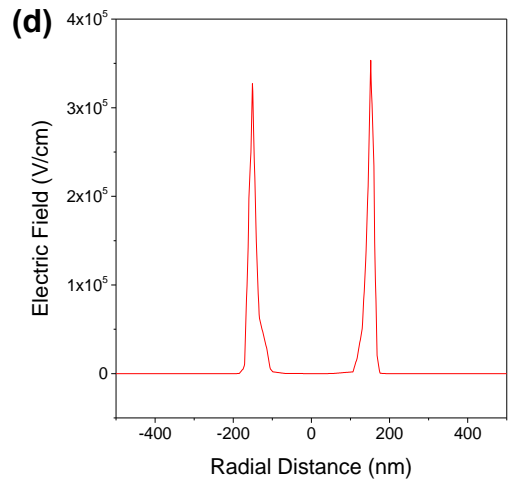

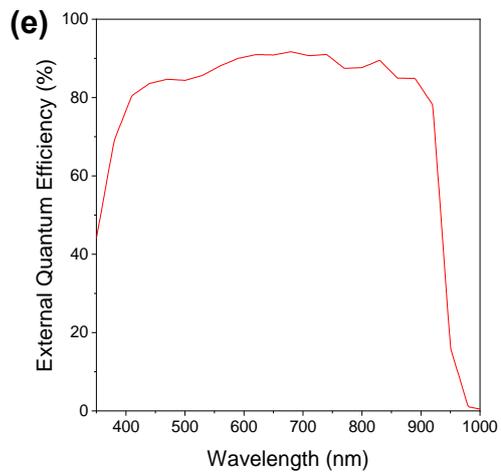

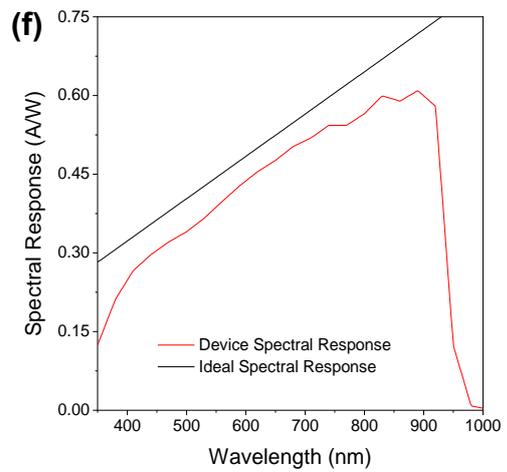

**Figure 2 | Simulation and quantum efficiency measurement of InP nanowire photodetector. a** Simulated absorption profile of the radial junction nanowire photodetector. For most wavelengths between 350-800 nm, the nanowire photodetector has near-unity absorption, leading to high sensitivity across a broad spectrum. **b** Simulated band alignment between n-ZnO and p-InP showing a type-II heterojunction with a large valence band offset. **c** and **d** shows simulated built-in electric field profile for nanowire radial junction at 0 V. A radial junction between heavily doped p-InP and AZO/ZnO leads to a built-in electric field of over $10^5$ V/cm at 0 V, leading to highly efficient charge carrier extraction. **e** External quantum efficiency measured at 0 V. **f** Photodetector spectral response measured at 0 V.

## 2.2. Single photon level detection

As indicated by the simulation, the InP/ZnO/AZO core-shell structure has a large internal build-in electrical field within a ~ 40 nm depletion region, enabling low dark current and highly efficient charge carrier separation of photo-generated electron and hole, which is ideal for achieving ultra-high photo sensitivity in photodetectors. To confirm this, we performed the photocurrent measurements at incident optical power as low as single-photon level. Figure 3a presents the I-V performance of the device under different incident optical powers. The dark currents are 23.7 pA at 0 V and ~ 9.85 nA under – 0.9 V, respectively. The extremely low dark current at 0 V suggests the high potential of the device to achieve high photosensitivity at 0 V, and the overall low leakage current under reverse bias also indicates the high device quality owing to high built-in electric field, low depletion region width, carrier selectivity, and passivation effect offered by core-shell heterojunction formed between AZO/ZnO and p-InP. With a further increase of the reverse bias, the current rises slowly until the voltage approaches breakdown voltage of around – 1.1 V, where the current increases drastically. It is worth noting that a small breakdown voltage is the direct evidence of an extremely high built-in electrical field formed in such a unique p-n nanowire structure.

To focus on self-powered operation, the photocurrent response at 0 V under different incident powers are shown in Figure 3b. The relation between photocurrent and incident power can be expressed by the equation:

$$I_{ph} \propto P^{\alpha} \qquad (2)$$

where $\alpha$ is the dimensionless exponent of the power law.[28] In Fig. 3b, a highly linear dependence between photocurrent and incident light power can be observed with an extracted α value of ~1.07, indicating excellent device behavior and defect-free material properties.

Meanwhile, this photodetector also exhibits a large dynamic range and is able to distinguish the photon flux with power down to the level of picowatts.

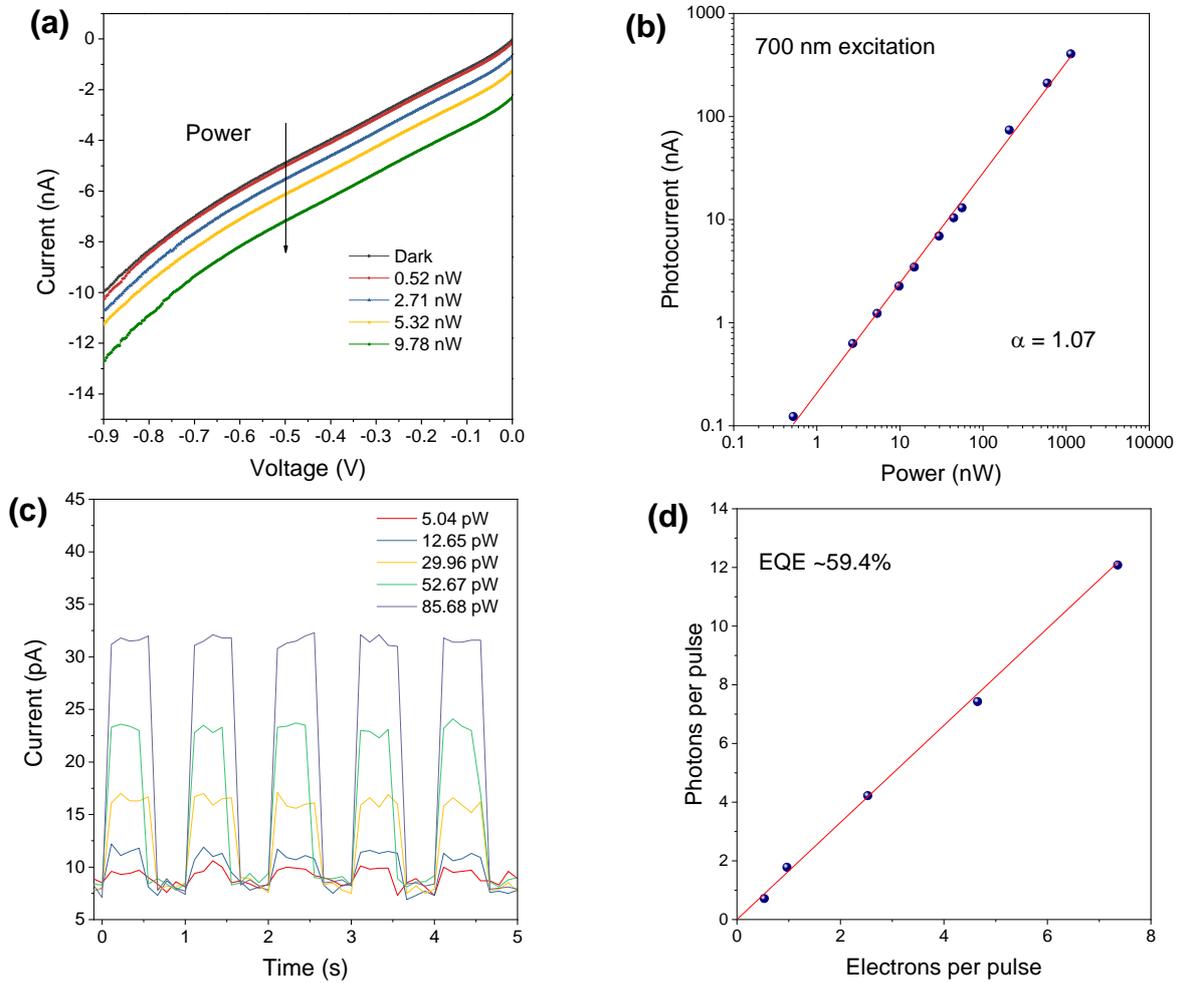

**Figure 3 | Photocurrent response under high power excitation condition and single photon level condition. a** Photocurrent measured at room temperature (300 K) in reverse bias under pulsed light excitation at 700 nm for different high excitation powers. **b** Photocurrent response shows a highly linear dynamic range from picowatts to microwatts. **c** Photocurrent measured at 300 K and 0 V bias for different ultra-low excitation powers. By attenuating 700 nm picosecond laser (6 ps pulse width) down to a single photon per pulse, the 5.04 pW is equivalent to the sub-single photon level. **d** Photocurrent response measured under pulsed excitation at 700 nm and 0 V at the single-photon level and at the room temperature, with an extracted quantum efficiency of 59.4%. It corresponds to more than 0.59 electrons per pulse under 1 single photon per pulse illumination condition.

To further investigate single-photon detection capability, the incident light power was attenuated down to the sub-single photon per pulse. The photocurrent can still be resolved from dark noise at room temperature, as presented in Figure 3c. The difference between dark and photocurrent at single photon per pulse is ~2 pA, which increases to ~24.5 pA when the number of photons is increased to 12 photons per pulse. In Figure 3d, we covert the incident light intensity and electrical current into photon per pulse and electron per pulse, respectively, to calculate the external quantum efficiency. About 0.59 electrons per pulse are generated with one photon per pulse illumination, corresponding to an external quantum efficiency of 59.4%.

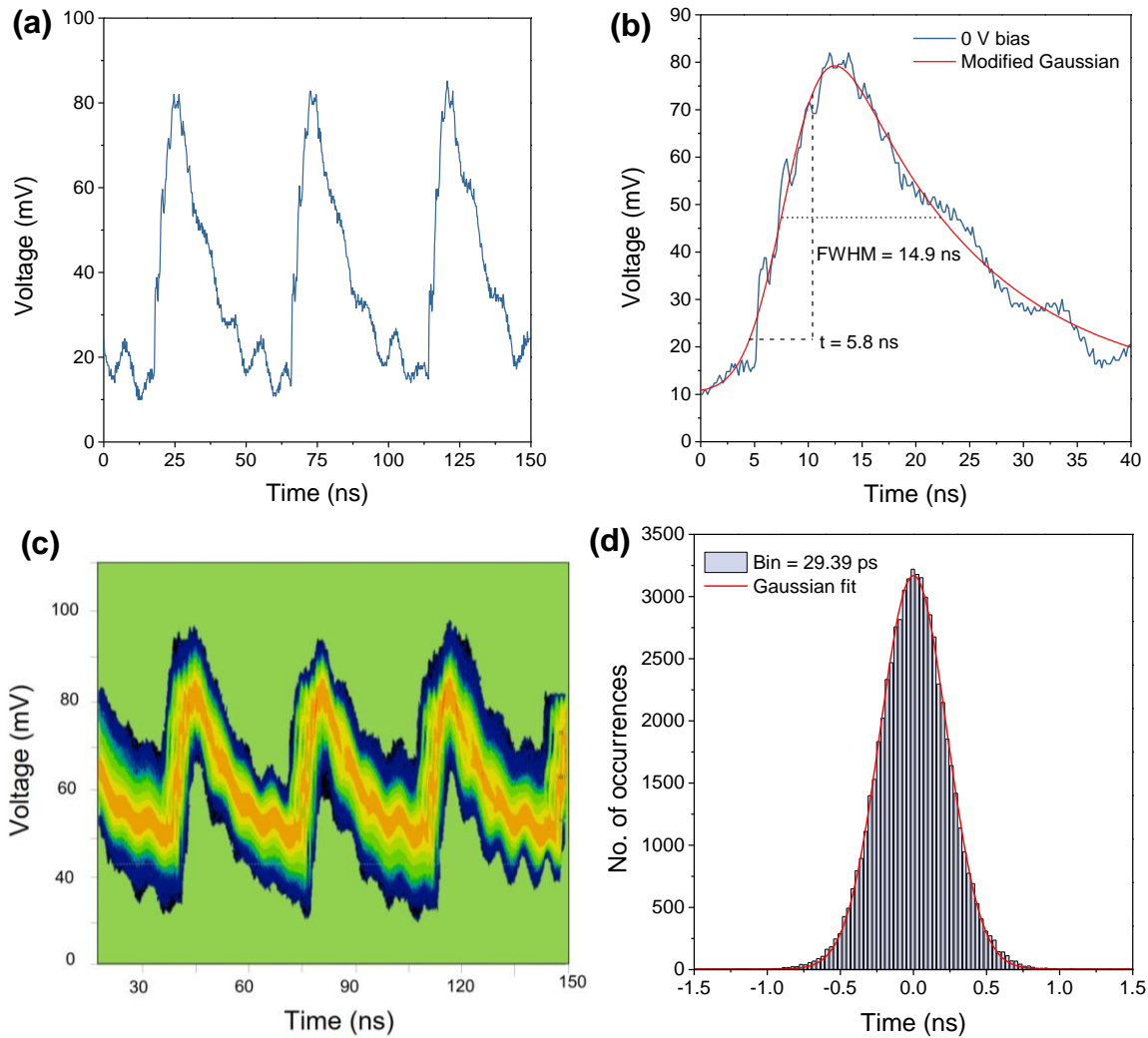

**Figure 4 | Temporal characteristics of the self-powered detector. a** Output pulse voltage from photodetector under 522 nm fs-laser excitation with a repetition rate of 20 MHz and power of 800 uW. **b** The single voltage pulse fitted with exponentially modified Gaussian function. The extracted rise time is 5.8 ns, and FWHM is ~14.9 ns. **c** Persistence plot of the photocurrent pulse waveform under 20 MHz 522 nm excitation at the power of 800 µW with 0 V bias. **d**

FWHM of the the histogram data, taken at 80% of the pulse rising edge from **c.** A Gaussian curve fitted to the data yields an estimated time jitter of 538 ps.

The small difference between the quantum efficiency calculated in Figures 2(e) and 3(d) may be due to more dominant Ohmic and recombination losses in low power regimes. This can, in the future, be tackled by reducing the active area of the device, better passivation (to reduce depletion region recombination), and/or by improving the sheet resistance of AZO. Nonetheless, the results successfully confirm that the device exhibits a high sensitivity down to the single photon level at room temperature, without an external bias.

*Temporal response of self-powered nanowire photodetector*

Apart from the sensitivity, the temporal response is another critical parameter of a photodetector. Figure 4a shows the photocurrent response plotted against time at 0 V under 522 nm pulsed laser excitation. As evident from the figure, the photocurrent appears periodically with a similar pulse shape, indicating the majority of wires has similar fast response time across the entire device. From the single signal pulse, a modified Gaussian function fits the curve well.[8] The 10 – 90 % rise time is estimated to be ~ 5.8 ns, and the FWHM is ~ 14.9 ns, suggesting the bandwidth of the photodetector has the potential to reach above 600 MHz. This rapid response time proves the device can be used as a self-powered high-speed photodetector.

Moreover, the speed of the photodetector can be further improved. In our device, there may be a trade-off between depletion region width and device speed. As emphasized in the manuscript, central to our device's single-photon level detection performance is a high built-in electric field with a short depletion region width to achieve extremely low dark current and efficient carrier collection. This, however, on the contrary, could slow down the device's time response due to the high junction capacitance (which is inversely proportion to the depletion width) and diffusion limited photocurrent arising from the large neutral (zero-field) region outside the small depletion region. In the future, device optimization may be performed through: 1) reducing the NW doping to slightly increase the depletion region width; and/or 2) decreasing the NW diameter to ensure the absorption mainly to occur in the depletion region; and/or 3) adding a thin dielectric layer between InP and ZnO to reduce the junction capacitance without significantly reducing the built-in electric field. Through a good balance among above key device parameters, we believe that improved time response with even higher detection efficiency can be achieved in our self-powered NW photodetector.

For applications in ultrafast optical communication and single photon counting, the timing jitter of the photodetector is another crucial criterion. The timing jitter is the uncertainty of timing information of captured arrival events due to the fluctuations of the output electrical signal. For photodetectors, the time from photon absorbed by semiconductor materials to the generated electrical current is not fixed owing to spatial variation of the electrical field and the finite transition time of the carriers within the device. To evaluate the timing jitter performance of the device, a mixed signal oscilloscope was used to collect the periodical output pulse traces at a fixed frequency of incident light excitation. Figure 4c shows the persistence of waveform trace records. Under this mode, the oscilloscope records every waveform that is triggered at the same trigger level. The hot color in the figure presents that there are more traces that overlap together. As expected, some waveforms do not lie within the general trend, showing a timing jitter in the device. A histogram representation of the number of occurrences of a specific type of waveform against time is plotted in Figure 4d. The plot of the histogram of the persistence data taken at 80% of the pulse rising edge was used to analyze the timing jitter. The FWHM extracted from the Gaussian fit to the histogram is 538 ps. This timing jitter originates from time variation in the time interval between the absorption of a photon and the generation of output electrical signals. The relatively large timing jitter in our device may be due to the non-optimized device speed and the non-uniformity of individual nanowires, which can be improved by further optimization of the device design and fabrication.

## 2.3. Discussion and future work

The self-powered photodetector presented in this paper is unique in several aspects, and we use this section to discuss the presented results, their implications, and future work. Firstly, central to the device's performance (i.e., sensitivity down to single-photon level) is a built-in electric field of more than $10^5$ V/cm at 0 V and a small depletion region of ~ 40 nm that leads to extremely low dark current and highly efficient charge carrier separation. In conventional, heavily doped p-n homojunction, such a high built-in electric field would form a tunnel junction with an Ohmic behavior, which will lead to limited or no photodetection due to a very high leakage current. On the contrary, in the current device, even with a high built-in electric field, carrier selectivity and a large valence band offset ensure minimum leakage of holes from InP to ZnO, thereby keeping the dark current sufficiently low to enable single-photon detection. A brief comparison between our device (operated under self-powered condition) and some commercial devices (measured under optimum bias voltage) has also been provided in Table S1 of the supporting information. Secondly, unlike most III-V devices, these devices do not rely on epitaxial growth, which in addition reduces the cost and complexity, and allows for flexibility in device design. For example, making a controlled radial p-n junction with high doping exceeding $10^{18}$ cm$^{-3}$ would be highly challenging, if not impossible, all by epitaxial growth. The use of a carrier selective contact to make a p-n instead of a lattice-matched epitaxial layer allows an additional level of freedom where several different kinds of carrier selective contacts can be tested in combination with non-epitaxial passivation layers for optimum performance. In addition to the electrical benefits, an optimized thickness of oxide layers also improves the absorption, making the device optically optimal, as well.

Furthermore, we would like to point out that the design and operation of such types of devices are not specific to InP and can be applied to other materials, thereby opening the possibility to fabricate self-powered, high-sensitivity photodetector for a broader wavelength regime beyond 1 μm. In particular, it would be interesting to explore the same device concept for Si and InGaAs, which are commercially more valuable than InP for photodetector because of their wide range of applications in silicon photonics and optical communications, respectively. In particular, the selective contacts for Si are way more advanced and well-studied for implementing such device design. In addition, Si can be more heavily doped than InP while maintaining very high quality with a carrier diffusion length exceeding several microns. Similarly, heavily doped $Ga_{0.47}In_{0.53}As$ can be grown lattice matched on InP to realize similar radial junction nanowire array photodetector through top-down etching. In the future, it is also worth exploring the fabrication of a similar device structure by deposition of carrier selective contact over high-quality epitaxially grown nanowires. Since the epitaxially grown nanowires have much lower surface defect density than that of the top-down etched nanowires, interface recombination at the p-n junction can be significantly reduced, leading to higher detector responsivity and improved temporal response.

Moreover, while MOCVD and MBE remain pivotal for the growth of complex structures (such as the ones consisting of different material compositions and doping levels), they suffer from high operational costs. We believe that a simple device structure such as the one reported in this work is expected to be more economical than epitaxially grown nanowire and planar photodetectors. For future application, a more detailed cost analysis is required to understand the cost *vs.* performance ratio of the proposed device in comparison to epitaxially grown III-V photodetectors.

For future work, the nearest aim would be to push the limit of sensitivity down to femtowatt levels by reducing the dark current below 100 fA. This is a realistic proposition considering currently, for 4 mm² area, we measure a current of 25 pA. With a further reduction of the device area to 100 μm x 100 μm the current can be reduced to less than 60 fA. Nonetheless, there would still be challenges in measuring such small power, where the stability of laser and electromagnetic noises from the environment can become an issue. Moreover, our device's speed and timing jitter can be improved through a thorough optimization of device design and fabrication.

## 3. Conclusion

In conclusion, we present high sensitivity radial p-n junction InP nanowire array photodetectors using carrier selective contact. The p-type InP nanowire arrays/n-type carrier selective contact (ZnO) device demonstrates the ability to detect single-photons at room temperature without an external bias, due to its high broadband absorption, extremely low dark current and efficient charge carrier separation enabled by a high internal built-in electrical field of more than $10^5$ V/cm at zero bias. Moreover, the detectors also exhibit a high bandwidth of above 600 MHz with a timing jitter of 538 ps. Finally, we discuss the future work to improve device performance, and how these devices stand out from conventional devices and open up possibilities for developing low-cost, high-sensitivity, self-powered photodetectors for future applications in remote sensing, space exploration, wireless environmental monitoring, and medical diagnostics.

## 4. Experimental Section

*Device fabrication:* The device fabrication started with InP nanowire etching using an ICP-RIE instrument (Samco RIE-400iP) at 180 °C using 1.5/25 sccm of $SiCl_4$/Ar gases at a chamber pressure of 0.1 Pa, with the RF/ICP powers of 50/200 W. The pattern for InP nanowire etching was generated using a Raith 150 Two Electron Beam Lithography (EBL) system, and $Cr/SiO_2$ is used as a mask. Next, the InP nanowires were planarized by spin coating diluted SU-8, and any surface coverage due to spin coating was removed using an oxygen plasma maintained at 100 W for 2 minutes. Next, ZnO followed by AZO was deposited in an ALD system. For all ALD deposition, the chamber pressure was maintained at 9 Pa, with the carrier gas ($N_2$) flow rate fixed at 150 sccm. At the same time, the pulse and purge times were fixed at 0.2 and 7 s, respectively. Next, the back contact (10 nm Zn/100 nm Au) was deposited in a sputter chamber and annealed at 400 °C in forming gas. Finally, 10 nm Ti/ 100 nm Au was deposited as the front contact.

*Photocurrent measurement:* Dark and photocurrent measurements were performed with a fiber coupled Fianium Supercontinuum laser and a Keysight B2902A precision source. A 99:1 fiber beam splitter was used to attenuate the laser intensity and the output laser power was calibrated by measuring the 99% fiber arm in real-time. The attenuated laser beam was sent into a customized microscope system where ambient light had been blocked. The current-voltage measurement was taken in a single sweep, with the sweep direction always from positive to negative voltage. The attenuated laser power in terms of photon per pulse is calculated as follows.

*Photon energy conversion:* The photon number can be estimated by the equation:

$$Photon\ number = \frac{P}{\frac{hc}{\lambda}} \times \frac{1}{f}$$

where $P$ is the excitation power, $f$ is the laser repetition rate, $\lambda$ is the incident light wavelength, $h$ is the Planck's constant and $c$ is light speed.

For the laser repetition rate of 20 MHz, the total photon energy amounts to 5.68 x $10^{-12}$ J/s, which is equivalent to 5.68 pW of laser power. In other words, 5.68 pW of laser power (@700 nm) corresponds to a single photon per pulse with a pulse repitition rate of 20 MHz ($20 \times 10^6$ per seconds).

**Photocurrent temporal measurement:** Temporal measurement was measured with Tektronix mixed signal oscilloscope 5 series and 1044/522 nm fs-laser, the frequency of which was doubled through a barium borate crystal. The measured device was directly connected to an oscilloscope with a 50 Ω input through an SMA cable without the application of any bias voltage.

## Acknowledgments

The authors would like to acknowledge the financial support from the Australian Research Council (ARC). Access to the fabrication facilities is made possible through the Australian National Fabrication Facility, ACT Node.

## Competing financial interests

The authors declare that they have no competing financial interests.